\documentclass{epl}
\usepackage{graphicx}  
\begin{document}
\title{On the role of mismatches in DNA denaturation}
\author{T. Garel and H. Orland}
\institute{Service de Physique Th\'eorique, CE-Saclay\\
91191 Gif-sur-Yvette Cedex, France}

\maketitle

\def\ie{{\it i.e.\/}~}
\def\eg{{\it e.g.\/}~}
\def\et{{\it et al.\/}}                                                         

\def\be{\begin{equation}}
\def\bea{\begin{eqnarray}}
\def\ee{\end{equation}} 
\def\eea{\end{eqnarray}}

\begin{abstract}
In the framework of the Poland Scheraga model of DNA denaturation, we
derive a recursion relation for the partition function of double
stranded DNA, allowing for mismatches between the two strands. 
This relation is studied numerically using standard parameters for the 
stacking energies and loop entropies. For
complementary strands (of length 1000), we find that mismatches are
significant only when the cooperativity parameter $\sigma$ is of order
one. Since $\sigma \sim O(10^{-5})$ in DNA, entropic gains from
mismatches are overwhelmed by the energetic cost of opening a
loop: mismatches are therefore irrelevant (and molecular recognition
of the strands is perfect). Generating random mutations with
probability $p$ on one strand, we find that large values of $\sigma$
are rather tolerant to mutations. For realistic (small) values of
$\sigma$, the two strands do not recombine, even for small mutation
fractions. Thus, molecular recognition is extremely selective.

\end{abstract}

\section{Introduction}
\label{introduc}
Natural DNA exists as a double-helix strand. Upon heating, the two
strands may separate. This unbinding transition is called DNA
denaturation. The reverse process of binding is called renaturation,
recombination, or, in a more biological wording, recognition.

The Poland-Scheraga model of DNA denaturation
\cite{Pol_Scher1,Pol_Scher2,Rich_Gut} has been recently
revisited from rather different points of view
\cite{Yer,Cau_Col_Gra,Ka_Mu_Pe,GMO,Ca_Or_St}.
The physics approach \cite{Cau_Col_Gra,Ka_Mu_Pe,GMO,Ca_Or_St} considers the phase
transition of the homopolymeric model (a  single pairing or stacking
energy). The transition is a first order transition, if the excluded
volume interaction between the two strands is properly  
taken into account. In a nutshell, one models the
weight of a loop of $l$ bases as $\sigma {e^{sl} \over {l}^c}$,
where $s$ is the entropy per base pair in a loop, $c$ is an exponent 
describing intra- and inter-strand excluded volume, and the
cooperativity parameter $\sigma$ is generally taken as unity. For
$c>2$, a first order (denaturation) transition occurs. Previous models 
relied on a value of $c \sim 1.8$, corresponding to a second order
transition.

On the other hand, Yeramian \cite{Yer} used the original
Poland-Scheraga model (with $c<2$) to study the denaturation
transition of real DNA sequences (that is with 10 different stacking
energies). Apart from minor differences (stacking vs. pairing
energies, inclusion of the entropy $s$ in the Boltzmann weight), this
realistic approach relies on the use of a factor $\sigma \simeq
10^{-5}-10^{-6}$, in marked contrast to the value $\sigma =O(1)$
used by physicists. 
The physical origin of $\sigma$ is the overlap of the $\pi$ orbitals
of the (deoxy)ribose rings of stacked bases. A small $\sigma$ value
means that opening a loop is very costly, and leads to a very sharp
(cooperative) first order like denaturation transition.
This sharpness allows Yeramian to identify coding regions in
real DNA sequences as
having a higher melting temperature than the non-coding ones.
The interplay between the values of $c>2$ and $\sigma$ was recently
studied \cite{Blo_Car} for some DNA sequences, in particular with
respect to the Meltsim program \cite{Meltsim}, where $c=1.75$ and
$\sigma \sim 10^{-5}$.

An important puzzle for physicists is how the extreme selectivity
required by the biological machinery can be achieved in spite of the
very high entropy of non selective binding. For instance, in a DNA
micro-array, single strands of DNA are grafted on a surface. When this
array is immersed in a solution containing complementary and
mutated strands, the recognition process occurs with a high accuracy,
with a very low rate of errors.

In this letter, we address these issues, allowing for
mismatches in the renaturation process. 
Our line of thought is as follows: if loops
are not costly energy-wise, one may a priory consider that mismatches
between the two strands (i.e. base $\alpha$ of the first strand being
paired or stacked with base $\beta$ of the second) play an important role.
In the homopolymeric case, it is easy to see that the sole effect of
mismatches is to replace the
exponent $c$ by $c-1$.

What we want to do in this letter is to adopt a pragmatic point of
view, leaving theoretical ideas for a
longer paper \cite{OG}. We will mostly work at $c=1.8$, using different
values of the cooperativity parameter $\sigma$. We have checked that
our results do not change significantly when using $c=2.15$ as
suggested in ref.(\cite{Ka_Mu_Pe}).

The plan of the paper is as
follows:  we first 
recall a few facts on the algorithmic aspects of the problems with-  and 
without- mismatches, and how to speed up calculations. The
comparison between the two cases will be first studied for
complementary sequences. For $\sigma$ small, mismatches
can be ignored, while they are essential for $\sigma \sim O(1)$.
Finally, we consider the case of non complementary sequences, and
argue that small values of $\sigma$ are not very tolerant to
mutations, in marked contrast to the case $\sigma \sim O(1)$. 

\section{Summary of Poland-Scheraga theory}
\label{Polscher}

We first consider two complementary strands of length $N$, and we denote by
$Z(\alpha)$ the partition function of the two strands originating at base $1$ and
paired at base $\alpha$. Using $Z(1)=1, Z(2)= e^{-\beta \varepsilon_{1,2;1,2}}$ as boundary conditions, 
the original Poland-Scheraga (PS) calculation can be recast in the
recursion relation 
\be
\label{partitps}
Z(\alpha)=Z(\alpha-1) \ e^{-\beta \varepsilon_{\alpha-1,\alpha;\alpha-1,\alpha}} +
\sigma \\
\sum_{\alpha^{\prime}=1}^{\alpha -2} Z(\alpha^{\prime}){\cal
N}(2(\alpha-\alpha^{\prime})) 
\ee
where $\varepsilon_{\alpha-1,\alpha;\alpha-1,\alpha}$ is the stacking
energy between bases $\alpha-1$ and $\alpha$, $\sigma$ is the
cooperativity parameter (that is the loop fugacity), and ${\cal
N}(2(\alpha-\alpha^{\prime}))$ is the number of closed loops of length
$2(\alpha-\alpha^{\prime})$. Equation (\ref{partitps}) expresses the
fact that pairing of base pair $\alpha$ 
may result either
from stacking  base pairs $\alpha-1$ and
$\alpha$ or 
closing
a loop at $\alpha'$.
The scaling form of the number of closed loops of length $l$ is
given by \cite{PGG} 

\be
\label{loops}
{\cal N}(l) = e^{s l} /l^c
\ee
where $s$ is the entropy per base in a loop, and $c$ is a critical
exponent, depending on the model used for the polymeric model. 
For a Brownian chain, one has $c=3/2$, whereas $c \simeq 1.8$ for a
self-avoiding chain. For interacting self-avoiding loops, it has been
argued that one should use $c \simeq 2.15$ \cite{Ka_Mu_Pe}.

Note that eq.(\ref{partitps}) can be backtracked (at least at low
temperature) in order to determine which fragments of the chain are
bound, and which ones are unbound. For that purpose, one iterates the
recursion (\ref{partitps}) till $\alpha =N$, keeping track of all the
$Z(\alpha)$. Going from $N$ to $N-1$, one keeps in
(\ref{partitps}) the one term of the r.h.s. 
which is dominant. If it is the first
term, it implies that the last pair is stacked, whereas if it is
one of the other terms of the sum, say $\alpha'$, there is a loop which ends 
at $\alpha'$. By iterating this
procedure, one ends up with the full conformational structure of the
chain.

We recall that the stacking energies have been shown to describe
nucleotides interactions in a much more accurate fashion than simple
base pairing. The idea is that in addition to the usual Crick-Watson
pairing, there is a big component of the binding energy originating
from the stacking of the ribose rings. In practice, this is modeled
by assuming that the interactions depend on pairs of adjacent
bases on the two strands, namely are of the form $\varepsilon_{i,i+1;j,j+1}$
instead of $\varepsilon_{i;j}$. Since there are four different bases, there
are in principle $4^4=256$ possible stacking energies, out of which
only 10 turn out to be non zero.

In the homogeneous case
$\varepsilon_{\alpha-1,\alpha;\alpha-1,\alpha}=\varepsilon$, a Laplace
transform with respect to $\alpha$ allows to solve exactly
equation (\ref{partitps}), yielding a continuous
(if $1<c<2$) or discontinuous (if $c>2$) transition. For ($c<1$), the
strands are bound at all temperatures. Note that in the homogeneous
case, the transition temperature is given by
\be
\label{tempcrit}
T_c={\varepsilon \over s - \log (1 + \sigma \sum_{l=1}^{\infty} 1 / l^c)}
\ee
This equation shows the very weak dependence of the critical
temperature on the loop exponent $c$ for physical values of $\sigma$
($\simeq 10^{-4} - 10^{-6}$).

From a more practical perspective, solving numerically equation
(\ref{partitps}) requires a CPU time of order $N^2$.
The Fixman-Freire method \cite{Fix_Fre,Yer3} reduces this CPU time to
order $N$ by approximating the loop factor by
\be
\label{Fix_Fre}
{1 \over {l}^c} \simeq \sum_{i=1}^I a_i \
e^{-b_i l}
\ee
In equation (\ref{Fix_Fre}) the number $I$ of couples $(a_i,b_i)$
depends on the desired accuracy. For a sequence of length $1000$, the
value $I=9$  gives an accuracy better than $0.5 \%$. Larger values
($I=14$) are
used in \cite{Yer} for lengths of 150000 base pairs or 
in the program Meltsim \cite{Meltsim}.

To be complete, a special treatment is required for
the extremities. Indeed, the number of conformations of a segment of
length $l$ at the extremities of the chain is
\be
{\cal N}({l}) \simeq e^{sl} l^{\gamma -1}
\ee
where the exponent $\gamma \simeq 1.15$. Thus, one should
slightly modify the above recursion at the origin to account for this
fact.

\section{Mismatch}

We again consider two strands of length $N$ and $N'$, and denote by $\alpha$
(resp. $\beta$) the base index of the first (resp. the second)
strand. Let $Z(\alpha,\beta)$ be the partition function with bases
$\alpha$ and $\beta$ paired (in these notations, the PS model
corresponds to $\alpha=\beta$). Using the same kind of 
boundary conditions as in
the previous Section, $Z(1,1)=1, Z(\alpha,1)=Z(1,\alpha)=0$, 
for any $\alpha = 2,...,N (N')$ and $Z(2,2)= e^{-\beta \varepsilon_{1,2;1,2}}$,
we may write
\bea
\label{partit1}
&&Z(\alpha,\beta)=Z(\alpha-1, \beta -1) \ e^{-\beta
\varepsilon_{\alpha-1,\alpha;\beta-1,\beta}}
\nonumber \\
&&+ \sigma 
\sum_{\alpha^{\prime}=1}^{\alpha -1} \sum_{\beta^{\prime}=1}^{\beta-1}
(1-\delta_{(\alpha^{\prime},\alpha-1)}
\delta_{(\beta^{\prime},\beta-1)})Z(\alpha^{\prime},\beta^{\prime}){\cal
N}(\alpha-\alpha^{\prime}+\beta-\beta^{\prime}) 
\eea

The interpretation of equation (\ref{partit1}) is the same as in
equation (\ref{partitps}). In the homogeneous case
($\varepsilon_{\alpha-1,\alpha;\beta-1,\beta}=\varepsilon$), one may
again solve the problem: the results are the same as the (PS) model,
except that $c$ is replaced by $c-1$, e.g. a discontinuous transition
occurs for $c-1>2$. 

This equation can be backtracked exactly as was explained in the previous 
section to obtain the conformations of the chains at low temperature.

In the following, we shall be interested in strands of equal lengths $N$.
From an algorithmic point of view, equation (\ref{partit1}) shows that
the calculation of the partition function, with our boundary conditions,
is of order $N^4$ (in the conclusion, we briefly discuss the influence of
the boundary conditions on this result). In what follows, we have used 
a Fixman-Freire approximation to reduce this $N^4$ power to a factor
$N^2\times I^2$.
Similar reductions were previously obtained
in studies of the denaturation of circular DNA \cite{Yer2}.

Defining
\be
S_j(\alpha, \beta) = e^{-b_j (\alpha + \beta)} \sum_{\alpha '=1}^{\alpha}\sum_{\beta
  '=1}^{\beta} e^{b_j (\alpha ' + \beta ')} Z(\alpha ', \beta ')
\ee
for $j=1,...,I$, it can easily be seen that these functions satisfy
the recurrence
\bea
&&S_j(\alpha, \beta) = e^{-b_j} (S_j(\alpha, \beta-1)+ S_j(\alpha-1 , \beta))-
e^{-2 b_j} S_j(\alpha-1, \beta-1) 
+ W_{\alpha \beta}
\nonumber \\
&+& 
V_{\alpha \beta} \biggl(S_j(\alpha-1 , \beta-1)
-e^{-b_j} (S_j(\alpha-1 , \beta-2)+S_j(\alpha-2 , \beta-1))
+ e^{-2 b_j} S_j(\alpha-2 ,\beta-2)\biggr) \nonumber \\
\eea 
where $W_{\alpha \beta}=
\sigma \sum_{l=1}^I a_l e^{- b_l}S_l ( \alpha-1,\beta-1)$ and
$V_{\alpha \beta}= e^{-\beta \varepsilon_{\alpha-1,\alpha;\beta-1,\beta}} - \sigma$.
The boundary conditions are
\be
S_j (1,1) =1,\  
S_j(\alpha,1)=S_j(1,\alpha)= e^{-b_j(\alpha-1)},\  
S_j(2,2)= e^{-2 b_j}+e^{-\beta \varepsilon_{1,2;1,2}}
\ee
for any $j=1,...,I$ and $\alpha=2,...,N$.

The partition function is given by
\be
Z(\alpha,\beta) = S_j(\alpha, \beta)-e^{-b_j}(S_j(\alpha-1,\beta)+S_j(\alpha,
\beta-1))+e^{-2 b_j} S_j(\alpha-1, \beta-1)
\ee

It is clear from these equations that the algorithmic complexity to
calculate the full partition function is $N^2 \times I^2$ instead of
$N^4$. With a number of components $I \simeq 10$, this allows to study
sequences of size up to $1000$ in a reasonable computer time.

Our numerical calculations have been done on the first
1000 bases of a fugu gene \cite{fugu}. We have
worked with $I=9$ ($a_i,b_i$) pairs. Since we
are interested in finite sequences, we took the Flory value $c=1.8$ for
the loop exponent (for $\sigma$ small, the melting curves for real DNA
sequences are very insensitive to the value of $c$, as can be inferred from eq.(\ref{tempcrit})). Finally the
stacking energies are the ones used in the program Meltsim \cite{Meltsim}.

The order parameter in DNA denaturation is usually taken as the
fraction of bound base pairs, denoted as $\theta$. This quantity can be measured by UV
absorption at 268 nm \cite{WB85}. 
The derivative $-d\theta / dT$ with respect to
temperature displays sharp peaks at the temperatures where various
fragments of the sequence open. In the following, for practical
reasons, we shall study the specific heat rather than the order
parameter. Indeed, for a homopolymer, the fraction $\theta$ is
proportional to the internal energy of the chain, and thus $-d\theta /
dT$ is proportional to the specific heat. In the case where there is a
non homogeneous sequence, it can be easily checked that the peaks of
the specific heat coincide with those of the order parameter derivative.

\section{Complementary strands: the role of $\sigma$}

Using equations (\ref{partitps}) and (\ref{partit1}), we have compared the
specific heat of the PS model with the specific heat of the model with
mismatches. 
 
First, in the homogeneous case (single stacking 
parameter) the two models differ, since the PS model undergoes a
continuous transition (with our choice $c=1.8$), whereas the model
with mismatches does not have a phase transition and
remains bound at all temperatures ($c-1=0.8<1$).
We have solved the recursion for a homopolymer of size 1000. For
$\sigma = 1$, the specific heats of the two models 
are wildly different, whereas for
$\sigma = 10^{-5}$ there is a sharp peak in both specific heats.
These peaks are
located exactly at the same place, the difference
between the two cases being that the height of the peak in the
mismatch case is finite, whereas it scales with the size of the chain
for the PS case.

\begin{figure}
\twofigures[height=4cm]{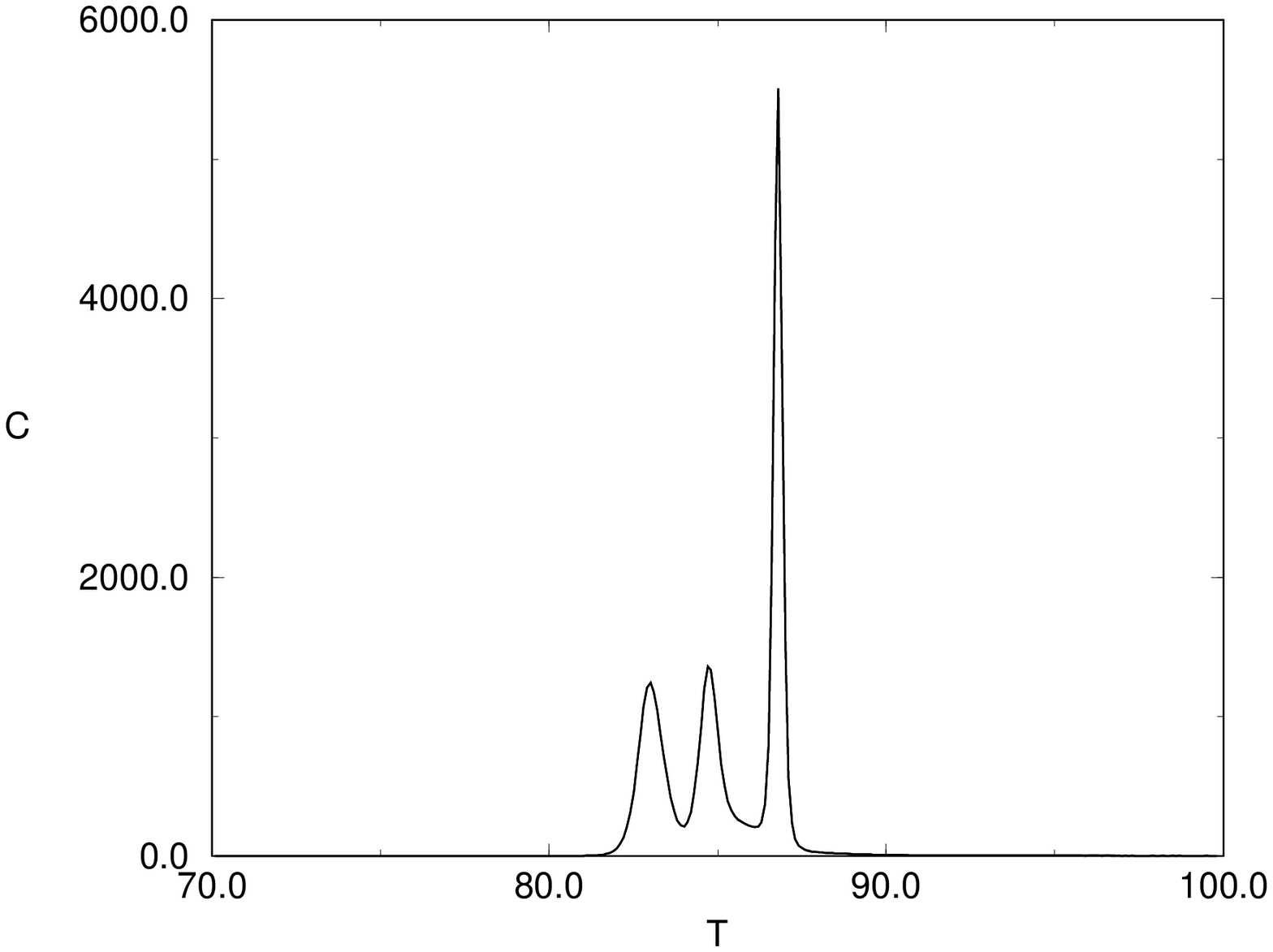}{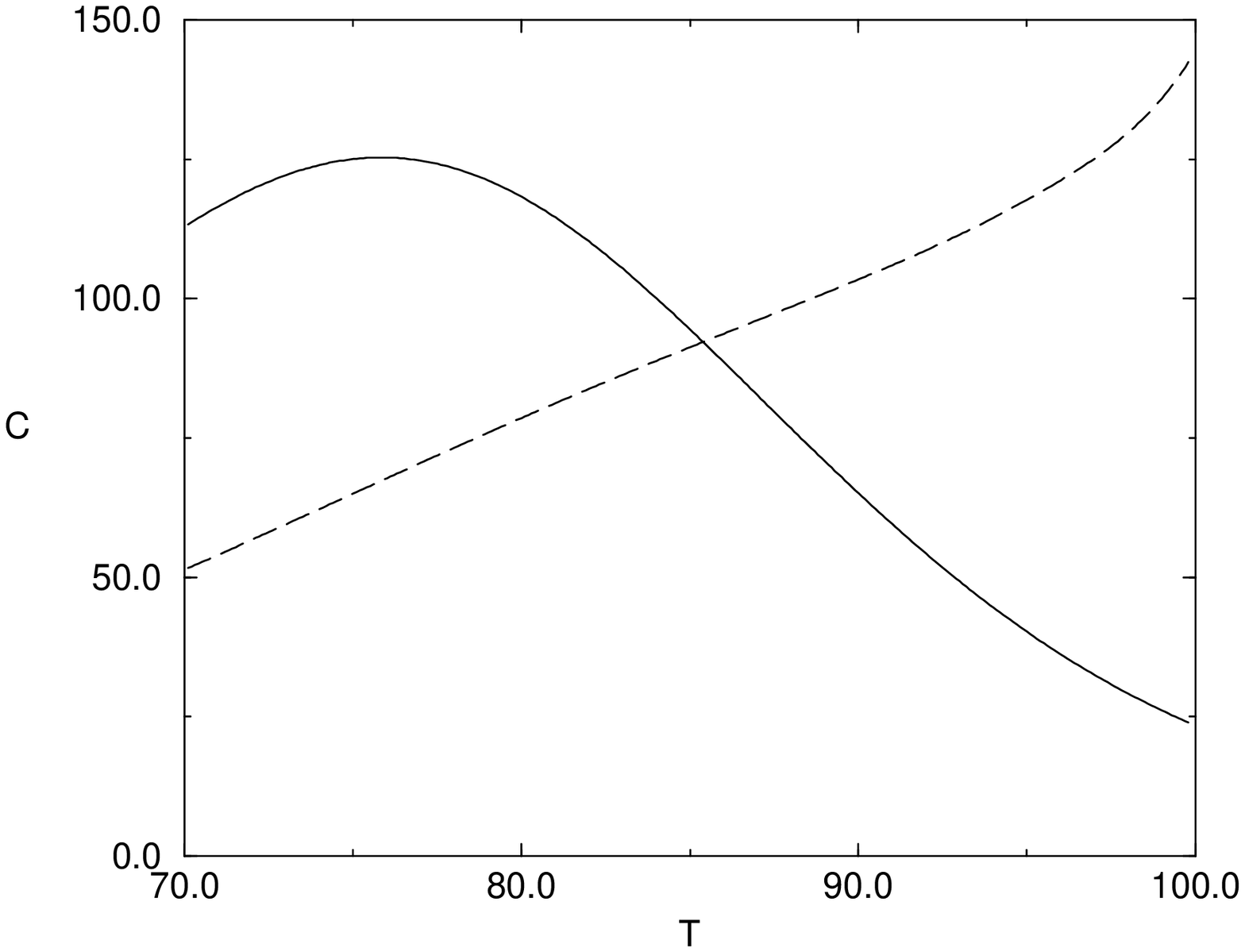}
\caption{Specific heat for $\sigma=10^{-5}$: the curves with and without
mismatches are undistinguishable}
\label{sigma.00001}
\caption{Specific heat with (continuous curve) and without
(long dashed curve) mismatches for $\sigma=1$}
\label{sigma1}
\end{figure}

We now consider the reference sequence defined above, for different
values of $\sigma$. For $\sigma = 10^{-5}$ (Fig. \ref{sigma.00001}), the curves can
be superimposed, whereas the two models differ considerably for
$\sigma=1$ (Fig. \ref{sigma1}). This means that for physical (small) values of
$\sigma$, mismatches between two complementary strands are negligible
at any temperatures. The physical reason is that although mismatches
are entropically favorable, they are suppressed by the high
energetical cost of initiating a loop. This cost vanishes for $\sigma
\simeq 1$.


\section{Random mutations: the role of $\sigma$}

We have generated random mutations on one strand, starting from the
complementary sequence. Our procedure is as follows: we mutate each
base on this strand independently with probability $p$; if the
mutation is accepted, the mutated base is different from the old one.
Equation (\ref{partit1}) is then implemented numerically.

\begin{figure}
\twofigures[height=4cm]{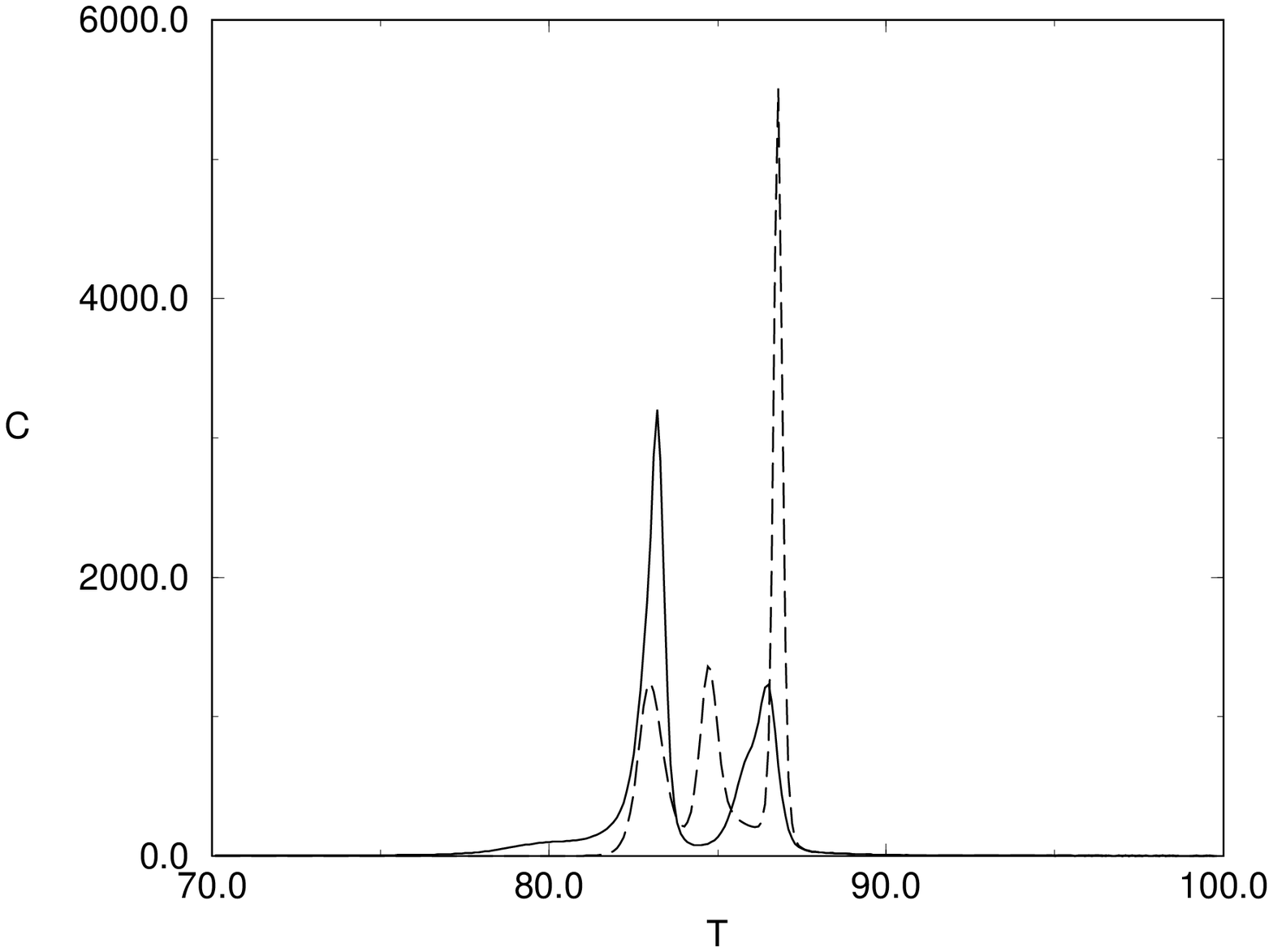}{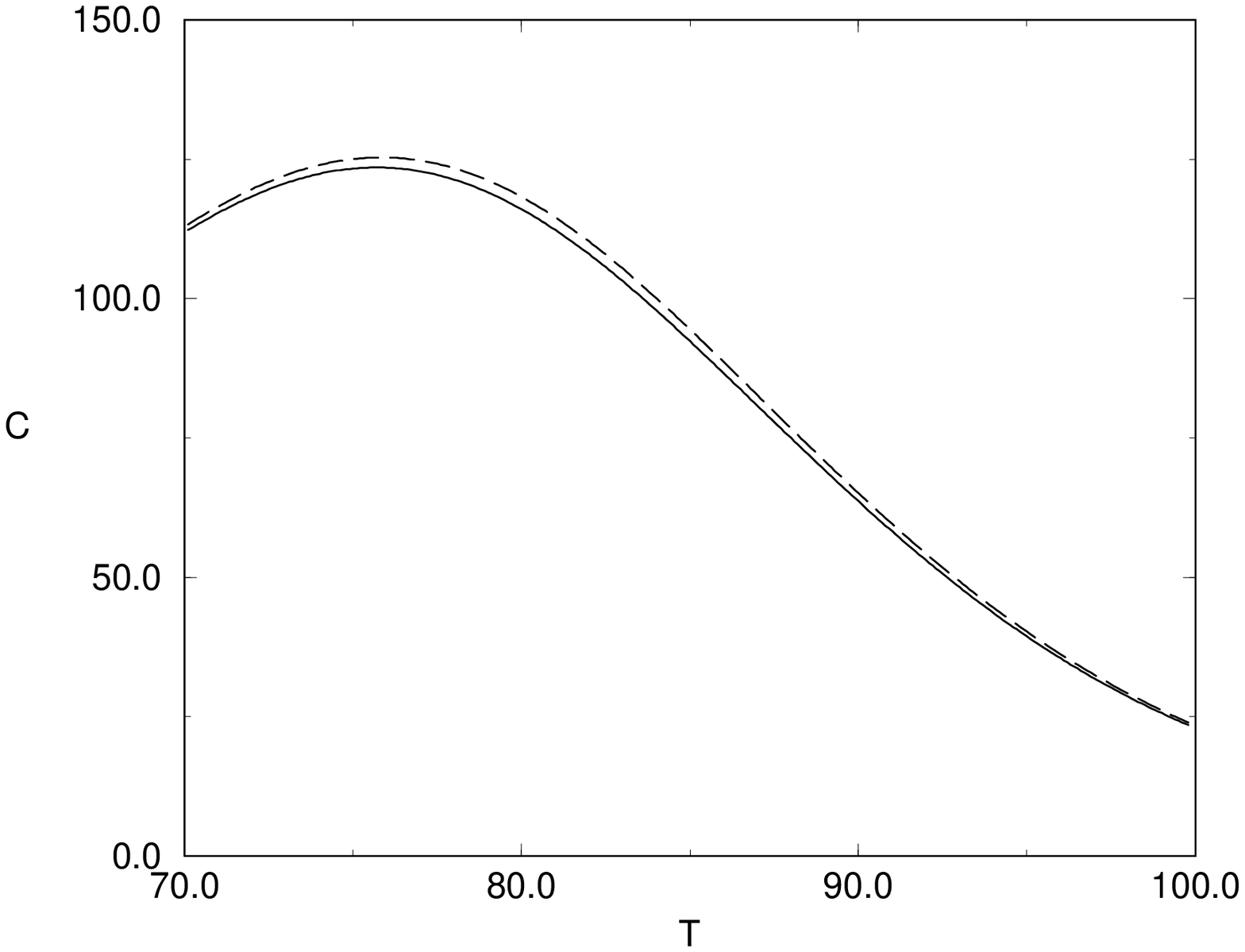}
\caption{Effect of mutations: Specific heat for $p=0.005$ (continuous
curve) and $p=0$ (long dashed curve) for $\sigma=10^{-5}$.}
\label{sigmamut.00001}
\caption{Effect of mutations: Specific heat for $p=0.005$ (continuous
curve) and $p=0$ (long dashed curve) for $\sigma=1$}
\label{sigmamut1}
\end{figure}

\begin{figure}
\onefigure[height=4cm]{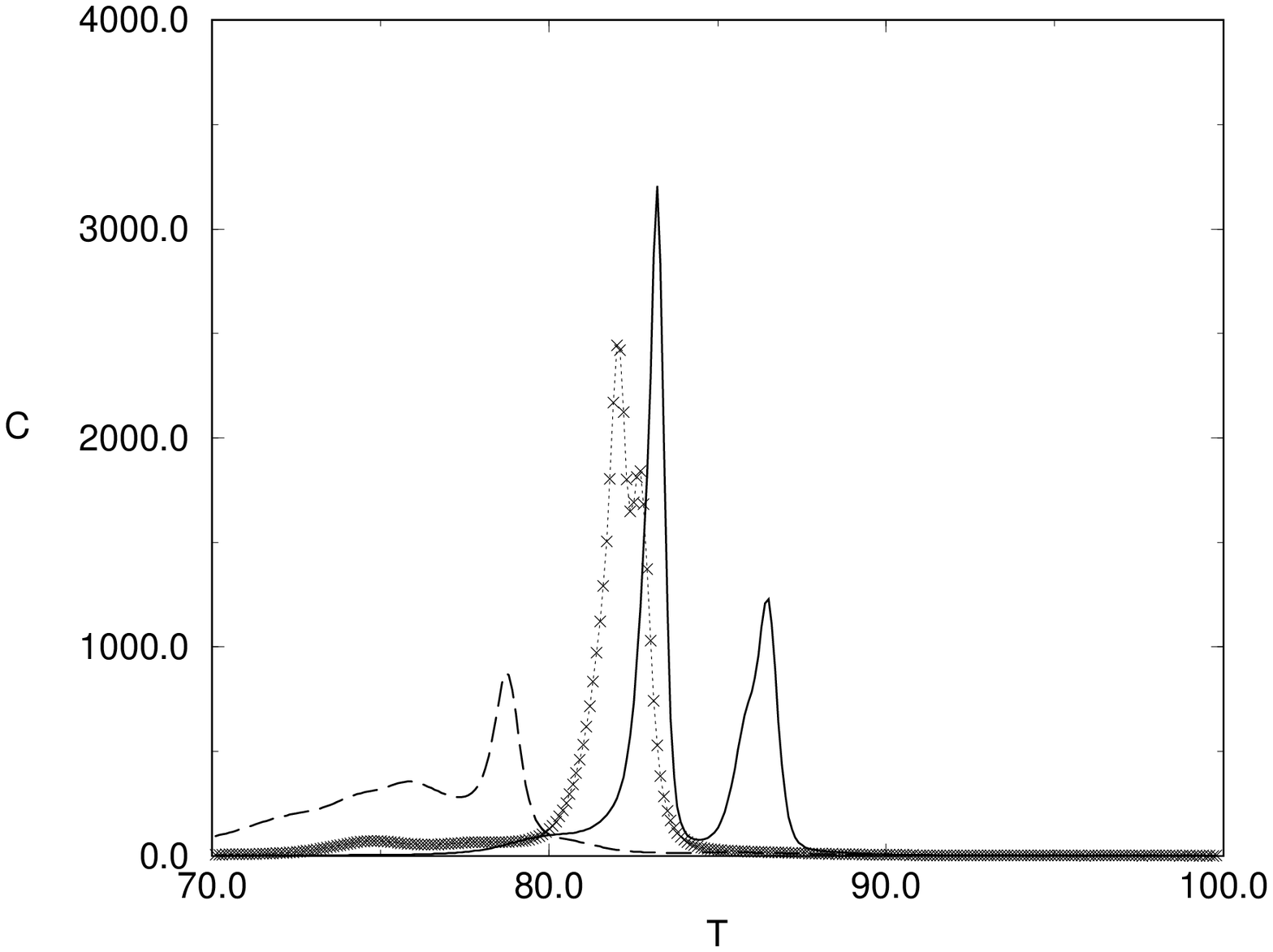}
\caption{Effect of mutations: Specific heat for $p=0.005$ (continuous
curve), $p=0.01$ (dotted curve with crosses) and $p=0.03$ (long dashed
curve), for $\sigma=10^{-5}$} 
\label{sigmamut.00001.p}
\end{figure}

For small values of $\sigma$, the melting temperatures decrease with
increasing values of $p$, as does the sharpness of the
peaks as shown in (Fig. \ref{sigmamut.00001} and 
\ref{sigmamut.00001.p}) for typical samples. Mutations
that are far along the sequence are disfavored and the strands
``unbind'' more easily. For a mutation rate $p$ of order 0.5\%
(corresponding  to a typical number of mutated bases between 4 and 6
in the reference sequence), recognition between the strands is already
poor. For larger values of $\sigma$ (Fig. \ref{sigmamut1}), the presence of
mismatches considerably lessens the role of mutations.

\section{Conclusion}
\label{Conclusion}
In this letter we have studied
why mismatches can be neglected in DNA denaturation and
recognition. We have argued that this is related mostly to the small
value of the cooperativity parameter $\sigma$.
We have seen that it is this very small value of $\sigma$ which forces
the very strong specificity in the DNA renaturation process.
In other words, $\sigma$
defines a length scale $l_L \sim 1/\sigma$ which characterizes the
typical distance between loops. For a realistic value of $\sigma \simeq
10^{-5}$, this amounts to lengths of order 100000 base pairs. One
expects significant effects of mismatches only beyond that length scale.

We have checked that all our results are very insensitive to the loop
exponent $c$, as long as the loop fugacity $\sigma$ is small enough.
Finally, the backtracking procedure can provide
physical informations about the boundaries of helical regions.

With our fixed
boundary conditions (where the two extremities of each strands are
fixed and paired) we have reduced the CPU time from $N^4$ to $N^2
\times I^2$ through the use of a Fixman-Freire approximation.
For general boundary conditions, we would get another $N^4$ factor from
relaxing the constraints on the extremities. 
This last factor can also be scaled
down. Indeed, consider adding  phantom paired links to the two extremities of 
each strands, namely $0_{\alpha}$,$0_{\beta}$ and,
$(N+1)_{\alpha}$,$(N+1)_{\beta}$. The stacking energy of these phantom
bases is taken to be 0. By this addition of phantom bases at the
origin and at the extremity of the two strands, we allow effectively
for open or closed boundary conditions on the two strands.
The error in doing so is to replace the true
$l^{\gamma -1}$ by $l^{-c}$ at the extremities of the chain. This 
simplification which affects only the extremal segments of the two strands
does not change the algorithmic complexity of the recursion
and the CPU time remains of order 
$N^2 \times I^2$. 

\section{Acknowledgements}
We thank E. Yeramian for discussions and for introducing us to the Fixman-Freire scheme.


\end{document}